\documentclass[final,1p,times]{elsarticle} 

\usepackage{amssymb}

\newcommand {\bc}{\begin{center}}
\newcommand {\ec}{\end{center}}
\def\lsim{\mathrel{\rlap{\lower4pt\hbox{$\sim$}}
    \raise1pt\hbox{$<$}}}               
\def\gsim{\mathrel{\rlap{\lower4pt\hbox{$\sim$}}
    \raise1pt\hbox{$>$}}}  
\newcommand {\bea}{\begin{eqnarray}}
\newcommand {\eea}{\end{eqnarray}}
\newcommand {\be}{\begin{equation}}
\newcommand {\ee}{\end{equation}}
\usepackage{graphicx}

\journal{Nuclear Physics A}

\begin{document}

\begin{frontmatter}



\title{Fermi liquid theory: A brief survey in memory of Gerald
E.~Brown}


\author{Thomas Sch\"afer}

\address{Department of Physics, North Carolina State
University, Raleigh, NC 27695}

\begin{abstract}
I present a brief review of Fermi liquid theory, and discuss recent 
work on Fermi liquid theory in dilute neutron matter and cold atomic 
gases. I argue that recent interest in transport properties of quantum 
fluids provides fresh support for Landau's approach to Fermi liquid 
theory, which is based on kinetic theory rather than effective field 
theory and the renormalization group. I also discuss work on non-Fermi 
liquids, in particular dense quark matter.
\end{abstract}




\end{frontmatter}


\section{Introduction}
\label{sec_intro}

 One of Gerry's main scientific pursuits was to understand the nuclear 
few and many-body problem in terms of microscopic theories based on 
the measured two and three-nucleon forces. One of the challenges of 
this program is to understand how the observed single-particle aspects 
of finite nuclei, in particular shell structure and the presence of 
excited levels which carry the quantum numbers of single particle states, 
can be reconciled with the strong nucleon-nucleon force, and how single 
particle states can coexist with collective modes. A natural framework
for addressing these questions is the Landau theory of Fermi liquids.
Landau Fermi liquid theory describes a, possibly strongly correlated, 
Fermi system which is adiabatically connected to a free Fermi gas. In 
particular, the system has a Fermi surface, and the excitations are 
quasi-particles with the quantum numbers of free fermions, but with 
modified dispersion relations and effective interactions. These 
quasi-particles coexist with collective modes, for example zero sound. 

 Gerry reviewed Fermi liquid theory in a number of his books and 
other writings. The conference that celebrated his 60'th birthday
was titled ``Windsurfing the Fermi Sea'' \cite{Kuo:1987jd}.
In the introduction of {\it Unified Theory of Nuclear 
Models and Forces} (3rd edition, 1970) Gerry writes:

\vspace*{0.3cm}
\bc\begin{minipage}{0.7\hsize}
{\it Many improvements could have been made, especially in 
Chapter XIII on effective forces in nuclei, but time is 
short, and I shall make them in later editions, when I am 
too old to ski. Of course, nobody will be interested in 
the subject by then.}
\end{minipage}\ec

\vspace*{0.3cm}
This prediction turned out to be incorrect. In his final decade
at Stony Brook Gerry trained and mentored a remarkable group of 
students who have helped to reinvigorate the study of effective 
forces in nuclei \cite{Bogner:2003wn,Bogner:2009bt}. 

\section{Landau Fermi liquid theory}
\label{sec_fl}

 Consider a cold Fermi system in which the low energy excitations 
are spin 1/2 quasi-particles. Landau proposed to define a distribution 
function $f_p=f_p^0+\delta f_p$ for the quasi-particles. Here, $f_p^0$
is the ground state distribution function, and $\delta f_p\ll f_p^0$
is a small correction. The energy density can be written as \cite{Baym:1991}
\be
 {\cal E} = {\cal E}_0 
  + \int d\Gamma_p\, \frac{\delta{\cal E}}{\delta f_p}\delta f_p
  + \frac{1}{2} \int\int d\Gamma_p d\Gamma_{p'}
    \frac{\delta^2{\cal E}}{\delta f_p\delta f_{p'}} 
    \delta f_p\delta f_{p'} + \ldots\, , 
\ee
with $d\Gamma_p=d^3p/(2\pi)^3$. Functional derivatives of ${\cal E}$ 
with respect to $f_p$ define the quasi-particle energy $E_p$ and the 
effective interaction $t_{pp'}$
\be 
\label{ep_flt}
E_p = \frac{\delta{\cal E}}{\delta f_p} \hspace{0.1\hsize}
t_{pp'}=\frac{\delta^2{\cal E}}{\delta f_p\delta f_{p'}} \, . 
\ee
Near the Fermi surface we can write $E_p=v_F(|\vec{p}|-p_F)$, where 
$v_F$ is the Fermi velocity, $p_F$ is the Fermi momentum, and $m^*=
p_F/v_F$ is the effective mass. We can decompose $t_{pp'}=F_{pp'}+G_{pp'}
\vec{\sigma}_1\cdot\vec{\sigma}_2$. On the Fermi surface the effective 
interaction is only a function of the scattering angle and we can 
expand the angular dependence as
\be 
F_{pp'} = \sum_l F_l\, P_l\left(\cos\theta_{\vec{p}\cdot\vec{p}'}\right) \, ,
\ee
where $P_l(x)$ is a Legendre polynomial, and $G_{pp'}$ can be expanded in 
an analogous fashion. The coefficients $F_l$ and $G_l$ are termed 
Landau parameters. 
 
The distribution function satisfies a Boltzmann equation
\be
\Big( \partial_t + \vec{v}_p\cdot\vec{\nabla}_x + 
    \vec{F}_p\cdot\vec{\nabla}_p\Big) f_p(x,t) = C[f_p]
\ee
where $\vec{v}_p=\vec{\nabla}_p E_p$ is the quasi-particle velocity, 
$\vec{F}_p=-\vec{\nabla}_x E_p$ is an effective force, and $C[f_p]$ 
is the collision term. Conserved currents can be defined in terms of
$f_p$ and the single particle properties $E_p$ and $v_p$. For example, 
we can write the mass density $\rho$ and mass current $\vec{\jmath}$ as
\be 
\rho =\int d\Gamma_p\, mf_p\, ,\hspace{0.3cm}
\vec{\jmath} =\int d\Gamma_p\, m\vec{v}_pf_p\, , 
\ee
where $d\Gamma_p=d^3p/(2\pi)^3$. The Boltzmann equation implies 
that the current is conserved, $\partial_0 \rho+\vec{\nabla}\cdot
\vec{\jmath}=0$. The conditions given in equ.~(\ref{ep_flt}) play 
an important is proving conservation laws for energy and momentum, 
and in establishing sum rules. 

 A different approach to Fermi liquid theory was popularized by 
Polchinski \cite{Polchinski:1992ed} and Shankar \cite{Shankar:1993pf}, 
see also \cite{Benfatto:1990zz,Salmhofer:1999uq}. 
Consider free non-relativistic quasi-particles near Fermi surface. The
system is described by the action
\be
\label{S_FLT}
S_{FL} = \int dt \int \frac{d^3p}{(2\pi)^3}
 \psi(p)^\dagger\left(i\partial_t-
      \left[\epsilon(p)-\epsilon_F\right]\right)\psi(p) 
\ee
Near the Fermi surface we can expand the momentum $\vec{p}=\vec{p}_F
+\vec{l}$ and
\be
 \epsilon(p)-\epsilon_F = \vec{v}_F(k)\cdot\vec{l}+O(l^2) 
\ee
We are interested in the question whether $S_{FL}$ is a possible 
fixed point of the renormalization group. For this purpose we 
study the scaling behavior of the action as $\vec{l}\to s\vec{l}$. 
The free fermion action is invariant under this rescaling provided
we assign the following scaling dimensions
\be 
\label{flt_scal}
 [k]=0,\hspace{0.05\hsize}
 [l]=1,\hspace{0.05\hsize}
 [\partial_t]=1,\hspace{0.05\hsize}
 [d^3p]=1,\hspace{0.05\hsize}
 [\psi]=-\frac{1}{2}\, . 
\ee
Consider now the effect of a four-fermion interaction
\be
\label{s_int}
  S_{int} =\int dt \left[ \prod_{i=1}^{4}\int
     \frac{d^3p_i}{(2\pi)^3}\right]
     \psi^\dagger(p_4)\psi^\dagger(p_3)
     \psi(p_2)\psi(p_1) \delta^3(p_{tot})U(p_1,p_2,p_3,p_4) \, .
\ee
For generic values of the momenta we can use the scaling dimensions
in equ.~(\ref{flt_scal}) to establish that interaction terms are 
irrelevant near the Fermi surface. An exception can occur if the 
dominant components of the momenta cancel and the delta function
constrains the small components $l_i$. There are two configurations
for which this occurs. One is BCS scattering $(\vec{p}_F,-\vec{p}_F)
\to (\vec{p}'_F,-\vec{p}'_F)$ which is characterized by  the 
interaction
\be
\label{BCS}
U(-\hat{p}_3,\hat{p}_3,-\hat{p}_1,\hat{p}_1) 
= V(\hat{p}_1\cdot\hat{p}_3) 
= \sum_l V_l P_l(\hat{p}_1\cdot\hat{p}_3)\, ,
\ee
where $P_l(x)$ are Legendre polynomials. The $V_l$ are marginal 
interactions. If any $V_l$ is attractive then loop corrections will 
grow logarithmically as $s\to 0$, and lead to BCS superfluidity. The 
other configuration is generalized forward scattering, where 
$\hat{p}_1\cdot\hat{p}_2=\hat{p}_3\cdot\hat{p}_4$, which is characterized 
by
\be
\label{ZS}
\left. U(\hat{p}_4,\hat{p}_3,\hat{p}_2,\hat{p}_1) 
\right|_{\hat{p}_1\cdot\hat{p}_2=\hat{p}_3\cdot\hat{p}_4}
= F\left(\hat{p}_1\cdot\hat{p}_2,\phi_{12,34}\right)
 +G\left(\hat{p}_1\cdot\hat{p}_2,\phi_{12,34}\right)
  \vec{\sigma}_1\cdot\vec{\sigma}_2\, .
\ee
We can write 
\be 
 F(z,0)=\sum_l F_l\, P_l(z) \, , 
\ee
and $F_l$ are the Landau Fermi liquid parameters defined above. The 
Fermi liquid parameters remain marginal even if loops are included. 
The fixed point characterized by equ.~(\ref{S_FLT}) and the parameters
$F_l$ is Landau Fermi liquid theory.

\begin{figure}[t]
\bc\includegraphics[width=0.44\hsize]{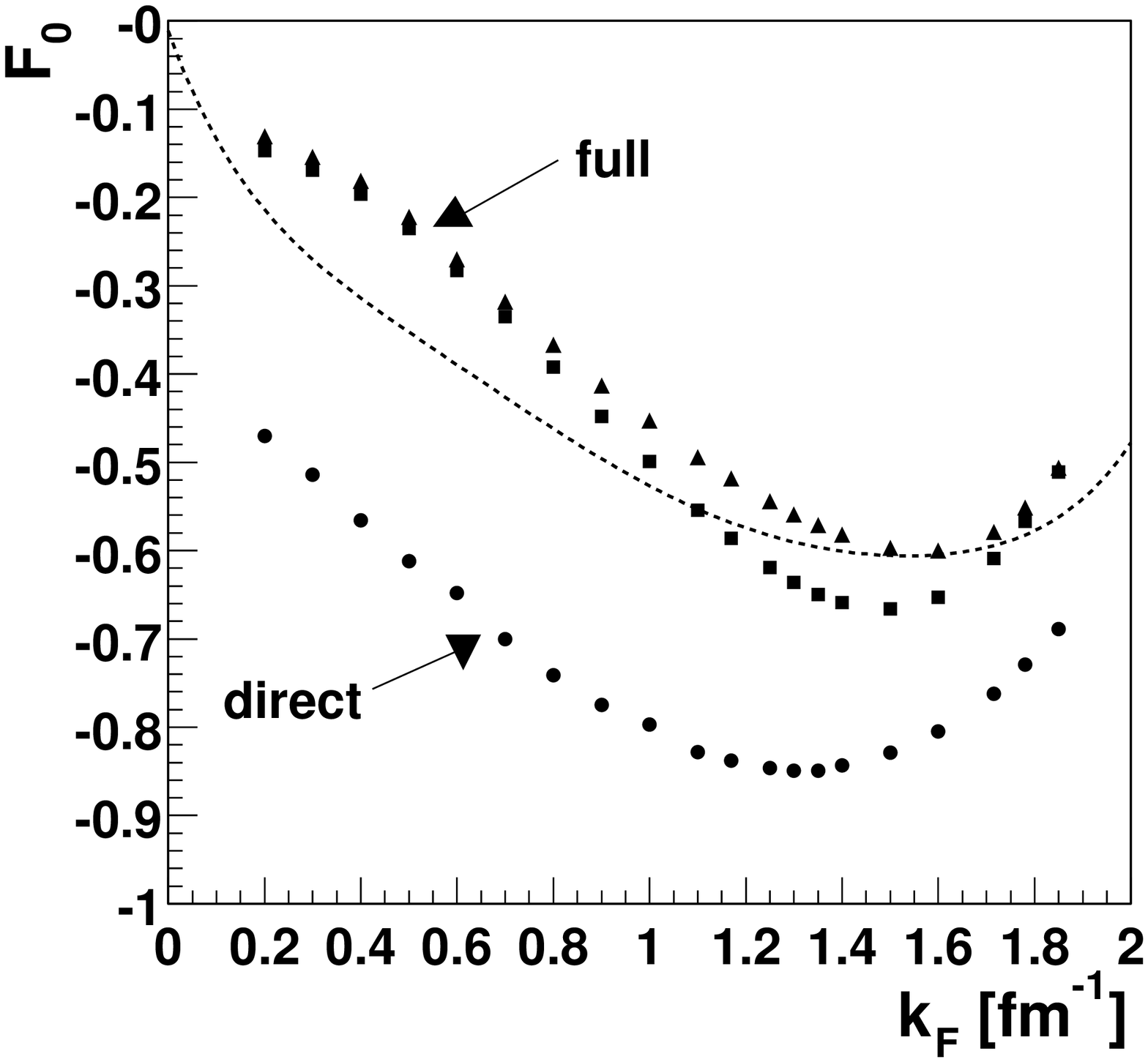}
\hspace{0.1\hsize}
\includegraphics[width=0.44\hsize]{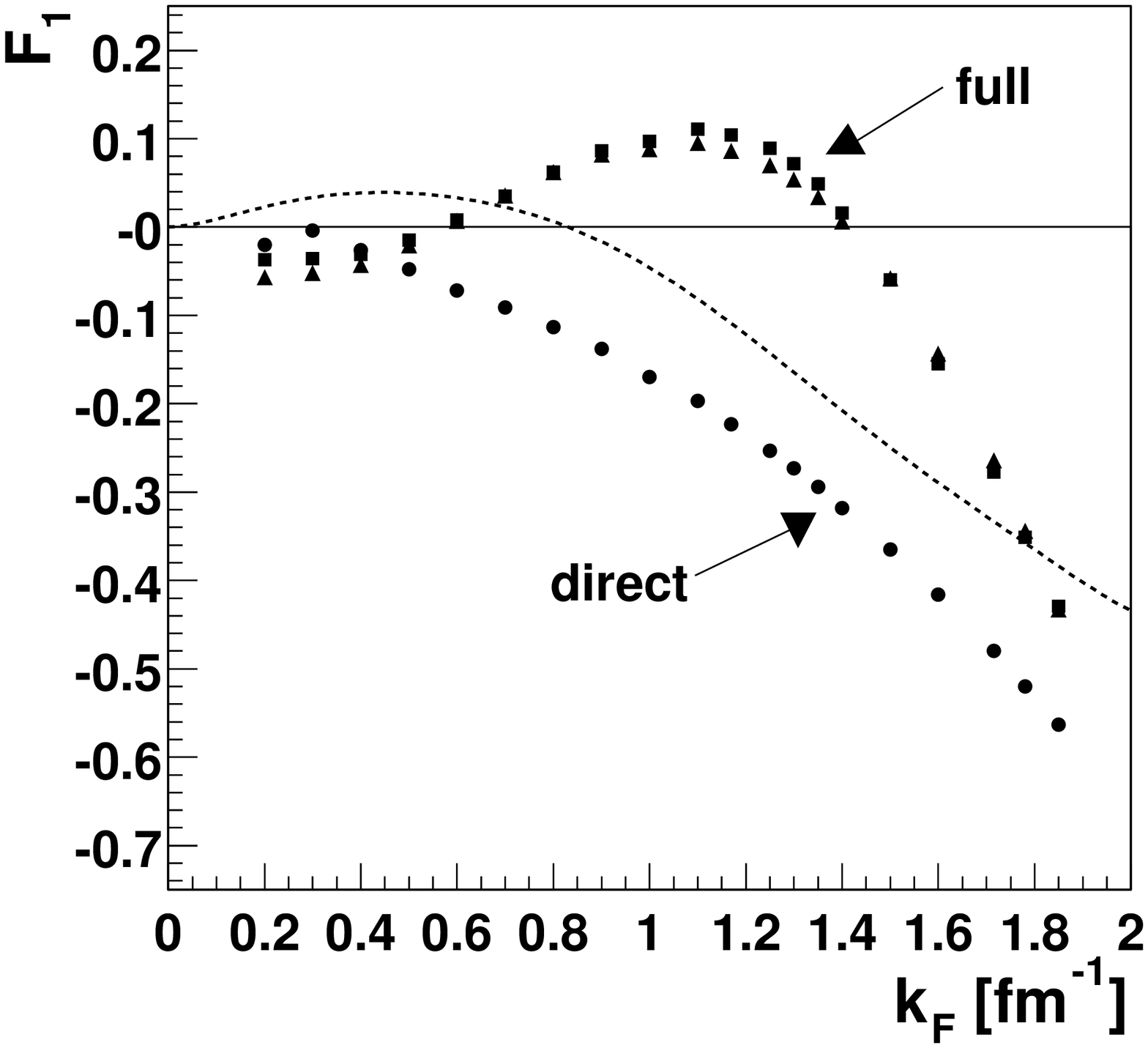}
\ec
\caption{\label{fig_fl}
Spin independent Landau parameters $F_0$ and $F_1$ in pure neutron 
matter as a function of the Fermi momentum $k_F$, from 
\cite{Schwenk:2001hg,Friman:2012ft}. The dots (labeled ``direct'')
corresponds to the bare interaction $V_{\rm low k}$, and were obtained 
with the wave function renormalization factor $Z=1$. The triangles 
and squares (labeled ``full') show the full results based on two
slightly different approximations to wave function renormalization 
factors, a static $Z$-factor in the case of the squares, and an 
adaptive $Z$-factor in the case of the triangles. The short dashed 
lines show the results of Wambach et al.~\cite{Wambach:1992ik}.}   
\end{figure}
 
\section{Neutron matter}
\label{sec_nm}

 In the case of neutron matter the renormalization group evolution
towards the Fermi surface was carried out explicitly by Schwenk, Friman,
and Brown \cite{Schwenk:2001hg}, see Fig.~\ref{fig_fl}. Schwenk et al.~use
$V_{\it low k}$ as the bare interaction at the UV scale $\Lambda=\sqrt{2}p_F$.
In nuclear matter three-nucleon forces are important, but there is no 
local $s$-wave interaction between three neutrons, and as a consequence
three nucleon forces are much less important in neutron matter. 
Schwenk et al.~employ the functional renormalization group equation 
in order to evolve the interaction to the Fermi surface. The resulting
Fermi liquid parameters $F_0$ and $F_1$ are shown in Fig.~\ref{fig_fl}. 
The parameter $F_1$ is related to the effective mass\footnote{This is 
curious relation from the point of view of effective field theory, as
it relates parameters of the EFT to a bare parameter. First of all, 
we note that both $m^*$ and $F_1$ are coefficients of marginal operators, 
so they can indeed occur together. Second, we can view this relation as 
a non-renormalization theorem that follows from Galilean invariance. 
Indeed, the bare mass is part of the Galilei algebra.}
\be 
\frac{m^*}{m} = 1 + \frac{F_1}{3}\, . 
\ee
We observe that in the regime 0.6 fm$^{-1}$ $<p_F<$ 1.4 fm$^{-1}$ 
the effective mass exceeds unity. It is interesting to compare 
this result to recent data from cold atomic Fermi gases at unitarity.
The unitary Fermi gas realizes the limit $a\to\infty$, where $a$ 
is the $s$-wave scattering length between the fermions. This system
is an interesting model system for dilute neutron matter, because
the $nn$ scattering length is anomalously large, $a_{nn}\simeq 
-19$ fm. Analyzing thermodynamic observables of the unitary Fermi 
gas just above the superfluid transition, Nascimbene et al.~ find
$m^*/m=1.13\pm 0.03$ \cite{Chevy:2009}, consistent with the sign of $F_1$ 
in Fig.~\ref{fig_fl}, but somewhat larger in magnitude.

\section{Non-Fermi liquid effective field theory}
\label{sec_nfl}

 Effective field theory ideas can also be applied to dense 
quark matter. As we will see, quark matter is not a Fermi liquid
in the strict sense, but many of the ideas of Fermi liquid theory
survive. In particular, quark matter above the critical temperature
for color superconductivity has quasi-particle and quasi-hole 
excitations that carry the quantum numbers of quarks. 

 QCD is a gauge theory, and the main new ingredient in an effective 
field theory of dense quark matter is the coupling of quarks to 
transverse gauge fields. This coupling is of the form $\vec{v}\cdot
\vec{A}$, where $\vec{v}$ is the velocity of the quark. At high density 
the Fermi momentum is large, and the emission of a low momentum gluon 
cannot change the velocity of the quark. An effective theory of 
quasi-quarks and quasi-holes interacting with soft gluons can be 
constructed by covering the Fermi surface with patches labeled by the 
local Fermi velocity\footnote{This idea is discussed in \cite{Shankar:1993pf}
as the basis of a ``large N'' approach to ordinary Fermi liquid theory.}
\cite{Hong:2000tn}. The effective lagrangian is given by 
\cite{Hong:2000tn,Schafer:2003jn,Schafer:2005mc}
\be
\label{l_hdet}
{\cal L} =\sum_v \psi_{v}^\dagger \left(iv\cdot D
   - \frac{1}{2p_F}D_\perp^2 \right) \psi_{v}
   + {\cal L}_{4f}
   + {\cal L}_{HDL} 
   -\frac{1}{4}G^a_{\mu\nu} G^a_{\mu\nu}+ \ldots ,
\ee
where $v_\mu=(1,\vec{v})$ if the four-velocity. The field $\psi_v$ describes 
particles and holes with momenta $p=\mu\,(0,\vec{v})+k$, where $k\ll\mu$. 
We can decompose $k=k_0+k_{\|}+k_\perp$ with $\vec{k}_{\|}=\vec{v}(\vec{k}\cdot 
\vec{v})$ and $\vec{k}_\perp = \vec{k}-\vec{k}_{\|}$. Hard gluon exchanges 
are described by a local four-fermion interaction ${\cal L}_{4f}$. This
interaction has the same structure as ordinary Fermi-liquid theory. 
Hard fermion loops are absorbed into the hard dense loop lagrangian
${\cal L}_{HDL}$. This interaction is somewhat subtle, because it cannot
be written in terms of a local lagrangian. Braaten and Pisarski showed
that \cite{Braaten:1991gm}
\be 
\label{S_hdl}
{\cal L}_{HDL} = -\frac{m^2}{2}\int\frac{d\hat{v}}{4\pi} 
  \,G^a_{\mu \alpha} \frac{v^\alpha v^\beta}{(v\cdot D)^2} 
G^b_{\mu\beta},
\ee
where $m^2=N_f g^2\mu^2/(4\pi^2)$ is the dynamical gluon mass and 
$\hat{v}$ is a unit vector. The hard dense loop action describes 
static screening of electric gauge fields and dynamic screening of 
magnetic modes. Since there is no static magnetic screening we
find that low energy gluon exchange is dominated by magnetic modes. 
The transverse gauge boson propagator is given by
\be
\label{d_trans}
D_{ij}(k) = \frac{\delta_{ij}-\hat{k}_i\hat{k}_j}{k_0^2-\vec{k}^2+
i\frac{\pi}{2}m^2 \frac{k_0}{|\vec{k}|}} ,
\ee
where we have assumed that $|k_0|<|\vec{k}|$. 

The gluon propagator is dominated by modes with momenta  $|\vec{k}|\sim 
(m^2 k_0)^{1/3}\gg k_0$. This relation leads to anomalous scaling relations
as we approach the Fermi surface. Consider a generic Feynman diagram and 
scale all energies by a factor $s$. From the relation quoted above we 
conclude that gluon momenta scale as $|\vec{k}|\sim s^{1/3}$, which 
implies that the momentum of a gluon is parametrically large to its
energy, and that gluons are far off-shell and space-like. 

\begin{figure}\bc
\includegraphics[width=5.cm]{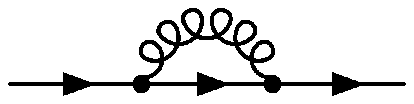}\hspace{-1.cm}
\includegraphics[width=5.cm]{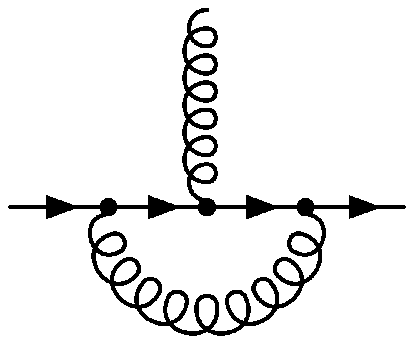}\hspace{-1.cm}
\includegraphics[width=5.cm]{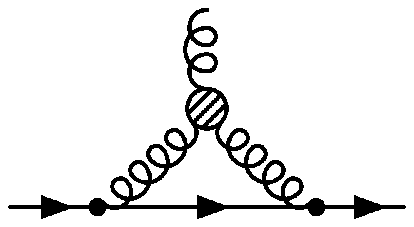}\ec
\caption{One-loop contributions to the quark self energy and the 
quark-gluon vertex. Near the Fermi surface the Feynman diagrams scale 
as $\omega\log(\omega)$, $\omega^{1/3}$ and $\omega^{2/3}$, respectively. }
\label{fig_mag}
\end{figure}

 The quark dispersion relation is $k_0\simeq k_{||}+k_\perp^2/(2p_F)$. 
For this relation to be consistent with the scaling of the gluon
momenta we must assume that gluon emission is accompanied by a 
transverse momentum kick. We find
\be 
k_0 \sim s, \hspace{0.5cm}
k_{||}\sim s^{2/3},\hspace{0.5cm}
k_\perp \sim s^{1/3},
\ee
and $k_0\ll k_{||}\ll k_\perp$. In this regime the quark and gluon
propagators are given by
\be
 S^{\alpha\beta}(p) = \frac{i\delta_{\alpha\beta}}
       {p_0 -  p_{||} - \frac{p_\perp^2}{2\mu}
              +i\epsilon \,{\mathrm{sgn}}(p_0)},
\hspace{0.75cm} 
 D_{ij}(k)  =  \frac{-i\delta_{ij}}
       {k_\perp^2-i\frac{\pi}{2}m^2\frac{k_0}{k_\perp}} \, , 
\ee
and the quark gluon vertex is $gv_i(\lambda^a/2)$. Higher order terms 
can be found by expanding the quark and gluon propagators as well as 
the HDL vertices in powers of the small parameter $\epsilon\!\equiv\!
\omega/m$.

 We can compare the scaling behavior of the interaction to the result 
for a Fermi liquid, see equ.~(\ref{flt_scal}). Consider the scale 
transformation $(x_0,x_{||},x_\perp) \to (s^{-1}x_0,s^{-2/3} 
x_{||},s^{-1/3}x_{\perp})$. The fields scale as $\psi\to s^{5/6}
\psi$ and $A_i  \to s^{5/6} A_i$. We find that the scaling dimension 
of all interaction terms is positive. The quark gluon vertex scales as 
$s^{1/6}$, the HDL three gluon vertex scales as $s^{1/2}$, 
and the four gluon vertex scales as $s$. Since higher order 
diagrams involve at least one pair of quark gluon vertices the expansion 
involves positive powers of $s^{1/3}$. The low energy theory 
is indeed weakly coupled, but the expansion involves fractional powers
(and logarithms) of the energy. 

\begin{figure}\bc
\includegraphics[width=4.5cm]{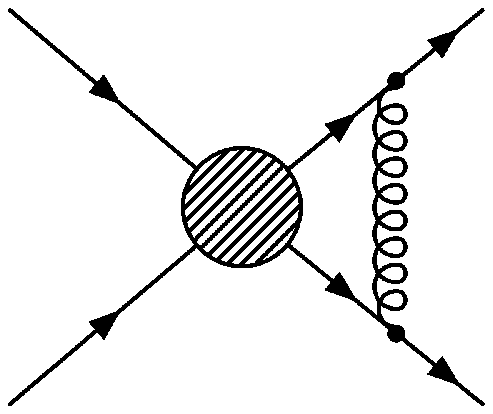}
\hspace*{1.5cm}
\includegraphics[width=4.5cm]{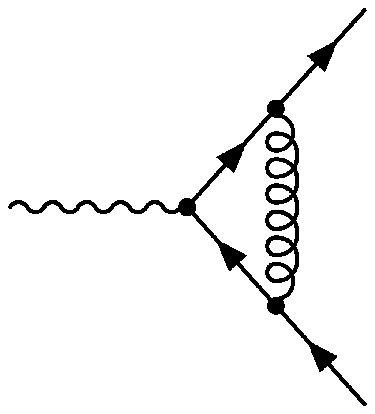}\ec
\caption{This figure shows the one-loop correction to the quark-quark 
interaction in the BCS channel (left panel) and vertex of an external 
current (right panel). Both diagrams are kinematically enhanced, and 
scale as $\log^2(\omega)$ and $\log(\omega)$, respectively. }
\label{fig_enh}
\end{figure}

 The effective field theory has been used to study a wide range 
of problems:

\begin{enumerate}

\item The quark self energy scales as $\Sigma(\omega)\sim \log(\omega)$,
see Fig.~\ref{fig_mag}. This result implies that the Fermi velocity and the 
wave function renormalization go to zero as $\omega\to 0$. The logarithm 
in $\Sigma(\omega)$  also leads to logarithmic terms in the specific heat 
\cite{Holstein:1973zz}. All of these effects are in contrast to a Fermi
liquid. In perturbation theory $\Sigma(\omega)\sim g^2 \log(\omega)$, 
and there are claims in the literature that perturbation theory breaks 
down at the scale $\omega\sim\exp(-1/g^2)$. This is not the case, as 
shown in \cite{Schafer:2004zf}. We also find a QCD version of Luttinger's 
theorem \cite{Luttinger:1960}: The quark density is given by the volume 
of the ``Fermi surface'', defined by the condition that the inverse 
quark propagator $S^{-1}(\omega\!=\! 0,p)$ changes sign 
\cite{Schafer:2006hx}. 

\item Corrections to the quark-gluon vertex are dominated by the 
abelian diagram in Fig.~\ref{fig_mag}. The scaling rules imply 
that 
\be 
\Gamma^a_\mu = gv_\mu (\lambda^a/2) \left(1+O(\epsilon^{1/3})\right).
\ee
This is a QCD version of Migdal's theorem, which states that 
the renormalization of the electron-phonon vertex is suppressed
by the ratio $\sqrt{m/M}$, where $m$ is the mass of the electron 
and $M$ is the mass of the ions \cite{Migdal:1958}. This factor
$\sqrt{m/M}$ is analogous to the small parameter in dense QCD, 
because it governs the ratio of the phonon velocity to the Fermi 
velocity of the electrons, which determines the ratio of the 
typical electron and phonon momenta. As a consequence of Migdal's
theorem we can compute the superfluid gap using an approximation
which includes loop correction to the gluon propagator, but does
include vertex corrections. 

\item In ordinary Fermi liquid theory, there are certain kinematical
regimes, the BCS and zero sound channels, in which the interaction 
is enhanced, see equ.~(\ref{BCS},\ref{ZS}). The same situation 
arises in non-Fermi liquid quark matter. Consider rescattering in 
the BCS channel as shown in Fig.~\ref{fig_enh}. The diagram scales
as $g^2\log^2(\omega)$, and the interaction becomes non-perturbative
at the scale $\omega\sim \exp(-1/g)$. This effect leads to color 
superconductivity, and a parametrically large gap $\Delta\sim 
\mu\exp(-1/g)$ \cite{Son:1998uk}. A similar phenomenon occurs when we 
consider the vertex of an external gauge field with coupling $e$.
The one-loop correction in the regime of small time-like momenta 
is  $\Gamma_\mu\sim eg^2v_\mu\log(\omega)$, and non-perturbative
effects become important for $\omega\sim \exp(-1/g^2)$ \cite{Brown:2000eh}. 

\end{enumerate}

\section{Transport theory}
\label{sec_transport}

 The examples discussed in the previous two sections are based
on the ``modern'', EFT based, approach to Fermi liquid theory. 
In this section I would like to argue that recent calculations
of transport properties of strongly correlated Fermi liquids
have reinvigorated Landau's original approach, which 
is based on kinetic theory. As an illustration I will discuss
recent work on the bulk viscosity of a dilute Fermi gas near 
unitarity\footnote{At unitarity the fluid is exactly scale
invariant, and the bulk viscosity vanishes.} \cite{Schaefer:2013oba}.
Interactions between quasi-particles play an essential role in 
this problem, because one can show that the bulk viscosity is zero 
if the ``interaction measure''
\be
 \Delta = {\cal E}-\frac{3}{2} P 
\ee
vanishes. In kinetic theory we view ${\cal E}$ as a functional 
of the distribution function $f_p$. The momentum density and the 
stress tensor are given by 
\bea 
\vec{\pi}\left(\vec{x},t\right) &=& \int d\Gamma_p\, \vec{p} 
  f_p \left(\vec{x},t\right)\, , \\
\label{pi_ij_kin}
\Pi^{ij}\left(\vec{x},t\right) &=& 
\int d\Gamma_p\, p^i v_p^j  
   f_p\left(\vec{x},t\right)
 + \delta^{ij} \left( \int d\Gamma_p \, E_p f_p\left(\vec{x},t\right)
   - {\cal E}\left(\vec{x},t\right)\right)\, .
\eea
Consistency with the laws of fluid dynamics requires that the 
momentum density is equal to the mass current, 
\be 
 \vec{\pi} \left(\vec{x},t\right) = \int d\Gamma_p\, m\vec{v}_p 
  f_p \left(\vec{x},t\right)\, ,
\ee
and that the momentum density is conserved, 
\be
 \partial_0 \pi^i \left(\vec{x},t\right)
 + \nabla^j\Pi^{ij}  \left(\vec{x},t\right) = 0\, . 
\ee
These relation can only be satisfied if $E_p=(\delta {\cal E})
/(\delta f_p)$, which is the fundamental relation in Landau's
approach to Fermi liquid theory. 

 In our work \cite{Schaefer:2013oba} we compute the equation of 
state and the quasi-particle properties as an expansion in the 
fugacity $z=\exp(\mu/T)$. This calculation is, strictly speaking, not 
a Landau Fermi liquid calculation because the fugacity expansion 
is reliable at high temperature, not at low temperature. However, 
since we employ a functional expansion of ${\cal E}$, the  calculation
can be generalized to the Fermi liquid case by replacing the bare 
interaction with an effective interaction, and by including the 
effects of quantum statistics in the collision term. Using the 
bare interaction the quasi-particle energy near unitarity is 
$E_p=\epsilon_p+{\it Re}\,\Sigma(p)+i{\it Im}\,\Sigma(p)$ with 
$\epsilon_p=p^2/(2m)$ and 
\bea
\label{Re_Sig}
{\it Re}\,\Sigma(p) &=& -\frac{4\sqrt{2}zT}{\sqrt{\pi}} 
       \frac{1}{a\sqrt{mT}}\sqrt{\frac{T}{\epsilon_p}} 
      F_D\left(\sqrt{\frac{\epsilon_p}{T}}\right)\, , \\
\label{Im_Sig}
{\it Im}\,\Sigma(p) &=& -
         \frac{2zT}{\sqrt{\pi}} 
         \sqrt{\frac{T}{\epsilon_p}} 
    {\it Erf}\left(\sqrt{\frac{\epsilon_p}{T}}\right)\, .  
\eea
where $F_D$ is Dawson's Integral, and ${\it Erf}$ is the error function. 
Because ${\it Im}\,\Sigma \sim zT \ll \epsilon_p\sim T$ quasi-particles
are well defined. 

 In an interacting fluid the total energy is the sum of a non-interacting 
term and the interaction energy. Bulk viscosity arises from the fact that
in an expanding system the relative magnitude of these two contributions 
may deviate from the equilibrium value. We find 
\be 
 \zeta = \frac{1}{96\pi^{5/2}}\, (mT)^{3/2} 
    \left( \frac{z\lambda}{a} \right)^2\, ,  
\ee
where $\lambda$ is the thermal de Broglie wave length. This result is 
consistent with the estimate $\zeta \sim \eta \Delta^2$, where $\Delta$
is the interaction measure, and $\eta\sim (mT)^{3/2}$ is the shear 
viscosity.

\section{Frontiers}
\label{sec_front}

 Fermi liquid theory continues to be a very active area of many-body
physics, and there are many new ideas that I cannot do justice to 
in a short article. One interesting area of research is to understand
the way anomalies in the underlying quantum field theory are implemented
in the Fermi liquid theory. Son and Yamamoto showed that anomalies 
give rise to non-zero Berry curvature of the Fermi surface \cite{Son:2012wh}.
The kinetic equation then leads to an anomalous conservation law, 
\be 
\partial_0 n + \vec{\nabla}\cdot\vec{\jmath} = 
  \pm \frac{1}{4\pi^2}\, \vec{E}\cdot\vec{B}\, , 
\ee
where the $\pm$ sign refers to left/right handed fermions, and $\vec{E}$, 
$\vec{B}$ are electric and magnetic $U(1)$ fields. Son and Yamamoto 
also showed how this equation arises in the high density effective field
theory described in Sect.~\ref{sec_nfl} \cite{Son:2012zy}.

 Another interesting area of research is the study of holographic 
Fermi and non-Fermi liquids \cite{Sachdev:2011wg,Faulkner:2010zz}.
These constructions are based on charged black holes embedded in 
asymptotically AdS (Anti de Sitter) spaces. According to the AdS/CFT
correspondence quantum gravity on this background is holographically
dual to certain field theories in flat space. By adding spinor fields
to the gravitational theory it is possible to realize holographic 
Fermi surfaces with a variety of Fermi liquid and non-Fermi liquid
spectral functions. Faulkner et al.~describe a holographic model
that gives a retarded spinor propagators of the form \cite{Faulkner:2010zz}.
\be
G_R(\omega,p) = \frac{h_1}{\omega-v_F(p-p_F)-\Sigma(\omega,p)}\, , 
\hspace{0.5cm}
\Sigma(\omega,p)= c\omega^{2\nu_{p_F}}\, , 
\ee
where $h_1,v_F,c$ are constants. The model can realize $\nu_{p_F}=
1/2$, corresponding to a marginal Fermi liquid such as the one 
described in Sect.~\ref{sec_nfl}, as well as  $\nu_{p_F}>1/2$ and
$\nu_{p_F}<1/2$. An important aspect of the AdS/CFT correspondence
is that it is fairly straightforward to compute transport properties
at strong coupling. As a result, one can relate quasi-particle 
properties, in particular the behavior of the fermion self energy 
near the Fermi surface, to transport properties, such as the 
conductivity spectral function.

\section{Final Remarks}

 I met Gerry in 1991 at a meeting in Peniscola, Spain. At the 
meeting, Gerry recruited me for long runs along the beach. I 
came to Stony Brook as a postdoc in 1992 and stayed until 1995. I
later returned as an Assistant Professor from 2000-2002. In the 
end, I only wrote one paper with Gerry \cite{Schafer:1993xt}, but 
the many lunch discussions, and the way Gerry attracted and mentored
his students and postdocs made a lasting impression on me. I will 
miss the unusual combination of his dry and sometimes harsh wit 
with the deep concern he showed for the welfare of those close to 
him. Gerry liked to talk about his ``eagles''\cite{Brown:2001xh}, 
Breit, Bethe, and Peierls. Just like these great scientists, Gerry 
appears to belong to a different era, one in which individual scientists 
could touch more people, and leave a more lasting impact than seems 
possible today.


Acknowledgments: This work was supported in parts by the US 
Department of Energy grant DE-FG02-03ER41260. 






\begin{thebibliography}{00}


\bibitem{Kuo:1987jd} 
T.~T.~S.~Kuo and J.~Speth,
``Windsurfing The Fermi Sea''. 
Proceedings, International Conference And Symposium On Unified 
Concepts Of Many Body Problems, Stony Brook, USA, September 4-6, 1986,
North Holland, Amsterdam (1987). 

\bibitem{Bogner:2003wn} 
S.~K.~Bogner, T.~T.~S.~Kuo and A.~Schwenk,
``Model independent low momentum nucleon interaction from phase shift 
equivalence,''
Phys.\ Rept.\  {\bf 386}, 1 (2003)
[nucl-th/0305035].

\bibitem{Bogner:2009bt} 
S.~K.~Bogner, R.~J.~Furnstahl and A.~Schwenk,
``From low-momentum interactions to nuclear structure,''
Prog.\ Part.\ Nucl.\ Phys.\  {\bf 65}, 94 (2010)
[arXiv:0912.3688 [nucl-th]].

\bibitem{Baym:1991}
G.~Baym, C.~Pethick, 
Landau Fermi Liquid Theory, 
Wiley, New York (1991).

\bibitem{Polchinski:1992ed}
J.~Polchinski,
Effective field theory and the Fermi surface,
Lectures presented at TASI 92, Boulder, CO,
hep-th/9210046.

\bibitem{Shankar:1993pf}
R.~Shankar,
``Renormalization group approach to interacting fermions,''
Rev.\ Mod.\ Phys.\  {\bf 66}, 129 (1994).

\bibitem{Benfatto:1990zz} 
G.~Benfatto and G.~Gallavotti,
``Renormalization-group approach to the theory of the Fermi surface,''
Phys.\ Rev.\ B {\bf 42}, 9967 (1990).

\bibitem{Salmhofer:1999uq} 
M.~Salmhofer,
``Renormalization: An introduction,''
Berlin, Germany: Springer (1999).

\bibitem{Schwenk:2001hg} 
A.~Schwenk, G.~E.~Brown and B.~Friman,
``Low momentum nucleon-nucleon interaction and Fermi liquid theory,''
Nucl.\ Phys.\ A {\bf 703}, 745 (2002)
[nucl-th/0109059].

\bibitem{Friman:2012ft} 
B.~Friman, K.~Hebeler and A.~Schwenk,
``Renormalization group and Fermi liquid theory for many-nucleon systems,''
Lect.\ Notes Phys.\  {\bf 852}, 245 (2012)
[arXiv:1201.2510 [nucl-th]].

\bibitem{Wambach:1992ik} 
J.~Wambach, T.~L.~Ainsworth and D.~Pines,
``Quasiparticle interactions in neutron matter for applications in neutron 
stars,''
Nucl.\ Phys.\ A {\bf 555}, 128 (1993).

\bibitem{Chevy:2009}
S.~Nascimbene, N.~Navon, K.~Jiang, F.~Chevy, C~Salomon,
``Exploring the Thermodynamics of a Universal Fermi Gas,''
Nature {\bf 463}, 1057 (2010)
[arXiv:0911.0747[cond-mat.quant-gas]].

\bibitem{Hong:2000tn}
D.~K.~Hong,
``An effective field theory of {QCD} at high density,''
Phys.\ Lett.\ B {\bf 473}, 118 (2000)
[hep-ph/981251].

\bibitem{Schafer:2003jn} 
T.~Sch\"afer,
``Hard loops, soft loops, and high density effective field theory,''
Nucl.\ Phys.\ A {\bf 728}, 251 (2003)
[hep-ph/0307074].

\bibitem{Schafer:2005mc}
T.~Sch{\"a}fer and K.~Schwenzer,
``Low energy dynamics in ultradegenerate QCD matter,''
Phys.\ Rev.\ Lett.\  {\bf 97}, 092301 (2006)
[hep-ph/0512309].

\bibitem{Braaten:1991gm}
E.~Braaten and R.~D.~Pisarski,
``Simple effective Lagrangian for hard thermal loops,''
Phys.\ Rev.\ D {\bf 45}, 1827 (1992).

\bibitem{Holstein:1973zz} 
T.~Holstein, R.~E.~Norton and P.~Pincus,
``de Haas-van Alphen Effect and the Specific Heat of an Electron Gas,''
Phys.\ Rev.\ B {\bf 8}, 2649 (1973).

\bibitem{Schafer:2004zf} 
T.~Sch\"afer and K.~Schwenzer,
``Non-Fermi liquid effects in QCD at high density,''
Phys.\ Rev.\ D {\bf 70}, 054007 (2004)
[hep-ph/0405053].

\bibitem{Luttinger:1960}
J.~M.~Luttinger, 
``Fermi surface and some equilibrium properties of a system of interacting
fermions'',
Phys.\ Rev.\ 119, 1153 (1960).

\bibitem{Schafer:2006hx} 
T.~Sch\"afer,
``Non-Fermi Liquid Effective Field Theory of Dense QCD Matter,''
Nucl.\ Phys.\ A {\bf 785}, 110 (2007)
[hep-ph/0608240].

\bibitem{Migdal:1958}
A.~B.~Migdal, 
Sov.\ Phys.\ JETP, 7 (1958) 996.

\bibitem{Son:1998uk} 
D.~T.~Son,
``Superconductivity by long range color magnetic interaction in high 
density quark matter,''
Phys.\ Rev.\ D {\bf 59}, 094019 (1999)
[hep-ph/9812287].

\bibitem{Brown:2000eh}
W.~E.~Brown, J.~T.~Liu and H.~c.~Ren,
``Non-Fermi liquid behavior, the BRST identity in the dense quark-gluon
plasma and color superconductivity,''
Phys.\ Rev.\ D {\bf 62}, 054013 (2000)
[hep-ph/0003199].

\bibitem{Schaefer:2013oba} 
T.~Sch\"afer and K.~Dusling,
``Bulk viscosity and conformal symmetry breaking in the dilute Fermi gas 
near unitarity,''
Phys.\ Rev.\ Lett.\ {\bf 111}, 120603 (2013)
[arXiv:1305.4688 [cond-mat.quant-gas]].

\bibitem{Son:2012wh} 
D.~T.~Son and N.~Yamamoto,
``Berry Curvature, Triangle Anomalies, and the Chiral Magnetic Effect 
in Fermi Liquids,''
Phys.\ Rev.\ Lett.\  {\bf 109}, 181602 (2012)
[arXiv:1203.2697 [cond-mat.mes-hall]].

\bibitem{Son:2012zy} 
D.~T.~Son and N.~Yamamoto,
``Kinetic theory with Berry curvature from quantum field theories,''
Phys.\ Rev.\ D {\bf 87}, no. 8, 085016 (2013)
[arXiv:1210.8158 [hep-th]].

\bibitem{Sachdev:2011wg} 
S.~Sachdev,
``What can gauge-gravity duality teach us about condensed matter physics?,''
Ann.\ Rev.\ Condensed Matter Phys.\  {\bf 3}, 9 (2012)
[arXiv:1108.1197 [cond-mat.str-el]].

\bibitem{Faulkner:2010zz} 
T.~Faulkner, N.~Iqbal, H.~Liu, J.~McGreevy and D.~Vegh,
``Strange metal transport realized by gauge/gravity duality,''
Science {\bf 329}, 1043 (2010).

\bibitem{Schafer:1993xt} 
T.~Sch\"afer, V.~Koch and G.~E.~Brown,
``Charge symmetry breaking and the neutron proton mass difference,''
Nucl.\ Phys.\ A {\bf 562}, 644 (1993).

\bibitem{Brown:2001xh} 
G.~E.~Brown,
``Fly with eagles,''
Ann.\ Rev.\ Nucl.\ Part.\ Sci.\  {\bf 51}, 1 (2001).

 \end{thebibliography}



\end{document}